\documentclass[conference]{IEEEtran}
\usepackage[utf8]{inputenc}
\pdfoutput=1

\usepackage[hyphens]{url}

\usepackage[dvipsnames]{xcolor}
\usepackage{colortbl}
\usepackage{amssymb}

\usepackage{pifont}
\newcommand{\xmark}{\ding{55}}%

\usepackage{breakurl}        
\usepackage[capitalise]{cleveref}

\usepackage{listings}
\definecolor{CodeGreen}{rgb}{0,0.5,0}
\definecolor{CodePurple}{rgb}{0.67,0.13,1}
\definecolor{CodeRed}{rgb}{0.73,0.13,0.13}
\newcommand{\snippet}[1]{\texttt{#1}}
\newcommand\textlcsc[1]{\textsc{\MakeLowercase{#1}}}

\begin{document}


\title{Vyper: A Security Comparison with Solidity Based on Common Vulnerabilities}
\author{
\IEEEauthorblockN{Mudabbir Kaleem}
\IEEEauthorblockA{University of Houston\\Houston, Texas}
\and
\IEEEauthorblockN{Anastasia Mavridou}
\IEEEauthorblockA{KBR / NASA Ames Research Center\\Mountain View, California}
\and
\IEEEauthorblockN{Aron Laszka}
\IEEEauthorblockA{University of Houston\\Houston, Texas}
}

\maketitle

\begin{center}
Published in the proceedings of the\\2nd Conference on Blockchain Research \& Applications\\for Innovative Networks and Services (BRAINS 2020).
\end{center}

\begin{abstract}
Vyper has been proposed as a new high-level language for Ethereum smart contract development due to numerous security vulnerabilities and attacks witnessed on contracts written in Solidity since the system's inception. 
Vyper aims to address these vulnerabilities by providing a language that focuses on simplicity, auditability and security. We present a survey where we study how well-known and commonly-encountered vulnerabilities in Solidity feature in Vyper's development environment. We analyze all such vulnerabilities individually and classify them into five groups based on their status in Vyper. 
To the best of our knowledge, our survey is the first attempt to study security vulnerabilities in Vyper. 
\end{abstract}

\section{Introduction}

Ethereum~\cite{buterin2013ethereum} is an open source, blockchain-based distributed computing platform that features smart contract~\cite{szabo1997formalizing} functionality. A smart contract is essentially a set of rules and procedures enforced digitally through pieces of Turing-complete code. Ethereum went live in July 2015, and is currently the most popular blockchain-based system for deploying smart contracts. Ethereum stores smart contracts in a public distributed database, i.e., a blockchain~\cite{nakamoto2008bitcoin}. Smart contracts are executed by the participants of the  network via transactions. Smart contracts exist on the Ethereum blockchain in opcode form. These opcodes are executed by the nodes of the Ethereum network inside the Ethereum Virtual Machine~\cite{wood2014ethereum} which is a stack based virtual execution environment. Having contracts executed inside the EVM ensures that the execution results are deterministic and identical in all the nodes regardless of their underlying computing capabilities.

Almost since the inception of Ethereum, Solidity~\cite{solidity} has been the most popular high level language for writing smart contracts. Developed by the contributors of the Ethereum project, Solidity continues to be the most popular tool for writing Ethereum smart contracts today. However, smart contracts written in Solidity are riddled with security vulnerabilities, which have been exploited in many highly publicized attacks on various Ethereum based projects~\cite{chen2019survey}. \cref{tab:incidents} lists some of the most notable incidents due to smart-contract vulnerabilities. Although Solidity has undergone many revisions and updates in order to address these vulnerabilities, multiple improvements can still be made. Perhaps the major reason of Solidity-written smart contracts encountering vulnerabilities, is that the language has been influenced by JavaScript and C++. Uninitiated developers, coming from web development and other backgrounds are lead to draw parallels between Solidity and these languages even where they do not exist. To be able to write secure smart contracts in Solidity, it is essential to have a sound understanding of the underlying Ethereum system and its various subtleties.

Recently, an alternate to Solidity, Vyper~\cite{vyper} has been developed to offer a better medium for writing smart contracts that are easier to understand. Vyper aims to make it harder for developers to intentionally write misleading or malicious code and also protects developers from unintendedly leaving vulnerabilities in their contract code. A stable version of Vyper is yet to be released but we offer a comparison between Vyper's latest beta version at the time of writing, i.e., 0.1.0-beta.15 and Solidity's latest release at the time of writing, i.e., v0.6.2. It is worth mentioning that Vyper does not claim to be a replacement for Solidity. It actually claims to strive towards goals of auditability, simplicity and security. To achieve these goals, it sacrifices various features and functionalities found in Solidity. This is done to make code written in Vyper more human readable, especially for the inexperienced users. If any of the more complex features of Solidity, which Vyper does not adapt, is required by the programmer then they will have to use Solidity for writing the smart contract. Moreover, Vyper also introduces additional features to support security and~readability.

\begin{table*}[t]
    \centering
    \caption{Recent incidents and cyber-attacks due to smart-contract vulnerabilities.}
    \label{tab:incidents}
    \begin{tabular}{| l | r | r | p{3.0cm} | p{2.9cm} | l |}
    \hline
    \bfseries Incident & \bfseries Date & \bfseries Amount & \bfseries  Vulnerabilities & \bfseries Mitigation & \bfseries References \\
    \hline\hline
King of Ether Throne &  February 6--8,  2016  &  98 Ether  &  Insufficient\,\, gas\,\, send, Exception for external call not handled  &  Manually sending back the failed transactions to participants  & \cite{king2016post} \\
\hline
Rubixi Vulnerability  &  April 2016  &  ~1.3k Ether  &  Wrong constructor name  & &  \cite{humiston2018smart} 
\\
\hline
GovernMental  &  April 2016  &  1.3k Ether  &  Insufficient gas  &  $\sim$50 ETH transaction fee paid to raise gas limit  & \cite{humiston2018smart} \\
\hline
The DAO Attack  &   June 16, 2016  &  ~3.6M Ether  &  Reentrancy  &  Addressed with a fork to the Ethereum blockchain  &  \cite{siegel2016understanding,finley2016million} \\
\hline
Parity Wallet Hack  &  July 19,  2017  &  ~150k Ether  &  Missing access control  & &  \cite{palladino2017parity} \\
\hline
Parity Wallet Freeze  &  November 6,  2017  &  $\sim$500k Ether  &  Unprotected suicide  & &  \cite{newman2017security} \\
\hline
POWH Coin Hack  &   January 28, 2018  &  ~2k Ether  &  Integer overflow  &  &  \cite{morisander2018biggest} \\
\hline
BEC Token Attack  &  April 2018  &  & Integer overflow &  & \cite{nvd2018cve} \\
\hline
Fomo3D Attack  &   August 22, 2018  &  ~10.5k Ether  &  Block stuffing &  & \cite{panda2018million} \\ 
\hline
SpankChain Attack  &  October 8, 2018  &  ~170 Ether  &  Reentrancy  &  & \cite{palmer2018spankchain}
\\
\hline
    \end{tabular}
\end{table*}

This paper presents a comparison of how various vulnerabilities, which are known to exist in the domain of smart contract development in Solidity, feature in Vyper's environment. The paper targets smart contract developers, users, and researchers who can use this resource to get up to speed with Vyper's current standing on known security issues. We are not currently aware of any other literature that provides such a comparison or details Vyper's vulnerabilities.  In \cref{sec:vyper}, the paper first outlines Vyper's principles and design goals along with the various features added in Vyper to improve upon Solidity. It also enumerates the features that exist in Solidity but have not been provided in Vyper in order to achieve its design goals. In \cref{sec:comp}, we present a taxonomy of the known vulnerabilities in smart contract development in Solidity and compare how Vyper features in each of those vulnerabilities.

\section{Vyper Principles and Features}
\label{sec:vyper}

Vyper, according to its official documentation, is a contract-oriented, pythonic programming language targeting the Ethe\-reum virtual machine. Although Vyper is still in development and a production ready release is yet to come out, we base this paper on the latest beta release, i.e., v0.1.0-beta.15. We do not expect there to be major design changes once a stable version comes out or once the Vyper project migrates from its current Python based compiler to a Rust based compiler~\cite{merriam2020update}. Vyper claims to be designed towards achieving the following three design goals or principles~\cite{vyper}:
\begin{itemize}
\item \textbf{Language and compiler simplicity:} Vyper aims to keep the language and the compiler implementation as simple as possible.
\item \textbf{Security:} Vyper aims to provide the programmer the ability to write smart contracts without any undesired vulnerabilities or loopholes.
\item \textbf{Auditability:} Vyper is aimed at making smart contracts easy to read for the users, especially those with insignificant prior experience with smart contracts or programming in general. Users should be able to identify malicious contracts with minimal effort. Vyper claims to give user readability preference over even the development experience for writing the contracts.
\end{itemize}

In order to achieve these desired goals, Vyper provides the following features not found in Solidity~\cite{vyper}:
\begin{enumerate}
\item	Bounds and overflow checking on array accesses and arithmetic;
\item	Support for signed integers and decimal fixed point numbers;
\item	Decidability: possible to compute the precise upper bound for the gas consumption of any Vyper function.;
\item	Strong typing, including support for units (e.g., timestamp, timedelta, seconds, wei, wei per second, meters per second squared);
\item	Small and understandable compiler code;
\item	Limited support for pure functions: anything marked constant is not allowed to change the state..
\end{enumerate}

Vyper also does not contain many of the features found in Solidity in order to achieve its desired objectives of minimal complexity and easy-to-do auditability by inexperienced developers. Following are the features that it does not contain:
\begin{enumerate}
\item	Binary Fixed Point: to avoid approximations associated with using binary fixed point.
\item	Recursive Calling: makes it impossible to compute an upper bound on gas consumption.
\item	Operator Overloading: makes writing misleading or complex code possible.
\item	Class Inheritance: makes understanding the code complex since precedence rules come into play in case of~conflicts.
\item	Inline Assembly: makes it impossible to search around for all instances of a variable name.
\item	Function Overloading: Can be confusing for the inexperienced programmer to keep track of which instance is~executed
\item	Infinite-Length Loops: makes it impossible to compute an upper bound on gas consumption.
\item	Modifiers: makes it easy to write misleading code.
\end{enumerate}

\section{Comparison of Vyper with Solidity's Vulnerabilities}
\label{sec:comp}

In this section we provide a detailed taxonomy of commonly known vulnerabilities in Solidity smart contracts and compare how each vulnerability fares in Vyper. Although numerous resources list the known vulnerabilities and attacks on smart contracts developed in Solidity, we use the vulnerabilities listed in Chen et al.~\cite{chen2019survey} as our base reference since we believe that this paper is the most comprehensive publication of these known vulnerabilities. Chen et al.\ attribute $19$ vulnerabilities in Ethereum systems security either to smart contract programming or to the Solidity language and toolchain. We analyze each of these vulnerabilities in Vyper's context and present our findings. Vyper may introduce additional vulnerabilities of its own in smart contract development that have not been observed in Solidity. However, given that the Vyper project is still in development and due to inadequate resources and test cases we do not attempt to study those in the present survey. This survey is intended to only explore how the currently known vulnerabilities translate to Vyper's environment.

We divide these $19$ known vulnerabilities into five groups (\cref{tab:comparison}) which are: 1)~vulnerabilities addressed by Vyper, 2)~vulnerabilities partially addressed by Vyper, 3)~vulnerabilities not addressed by Vyper, 4)~vulnerabilities that have already been mitigated in Solidity and 5)~vulnerabilities which we believe are not addressable by the programming language or the tool chain, despite being listed as such by Chen et al.

\begin{table}[t]
    \caption{Smart contract vulnerabilities in Vyper and Solidity. \textlcsc{\textbf{FA}, \textbf{PA}, and \textbf{NA}  stand for `Fully Addressed', `Partially Addressed', and `Not Addressed' by Vyper, respectively. Similarly, \textbf{AA} and \textbf{NP} stand for `Already Addressed by Solidity/Vyper' and `Not Addressable by language or toolchain,' respectively.}}
    \centering    \begin{tabular}{|{l}||{c}|{c}|{c}|{c}|{c}|}
    \hline
    \bfseries Vulnerabilities in Solidity & \bfseries FA & \bfseries PA & \bfseries NA  & \bfseries AA & \bfseries NP\\
    \hline
    \hline
    Integer overflow and underflow & \xmark &&&&\\
    \hline
    DoS with unbounded operation & \xmark &&&&\\
    \hline
    Unchecked call return value & \xmark &&&&\\
    \hline
    Reentrancy &&\xmark &&&\\
    \hline
    Delegatecall injection &&&\xmark &&\\
    \hline
    Forced Ether to contract &&&\xmark &&\\
    \hline
    DoS with unexpected revert &&&\xmark &&\\
    \hline
    Erroneous visibility &&&&\xmark &\\
    \hline
    Uninitialized storage pointer &&&&\xmark &\\
    \hline
    Erroneous constructor name &&&&\xmark &\\
    \hline
    Upgradeable contract &&&&&\xmark \\
    \hline
    Type casts &&&&&\xmark \\
    \hline
    Insufficient signature information &&&&&\xmark \\
    \hline
    Frozen Ether &&&&&\xmark \\
    \hline
    Authentication through \texttt{tx.origin} &&&&&\xmark \\
    \hline
    Unprotected suicide &&&&&\xmark \\
    \hline
    Leaking Ether to arbitrary address &&&&&\xmark \\
    \hline
    Secrecy failure &&&&&\xmark \\
    \hline
    Outdated compiler version &&&&&\xmark \\
    \hline
    \end{tabular}
    \label{tab:comparison}
\end{table}

The first group consists of currently existing vulnerabilities in Solidity that have been addressed in Vyper by providing an additional function or feature or disallowing specific features. These vulnerabilities can be completely avoided in Vyper. The second group consists of vulnerabilities which have been partially addressed by Vyper but still may exist if proper development practices are not followed by the developer. The third group consists of vulnerabilities that still exist both in Solidity and Vyper if the best programming practices and recommendations are not followed. The fourth group consists of historical vulnerabilities in the Solidity environment that were mitigated through later Solidity releases and do not exist in Solidity anymore and are also not present in Vyper. The fifth group consists of those vulnerabilities that have been listed by Chen et al~\cite{chen2019survey} as being caused by smart contract programming or the underlying Solidity language and toolchain. However, in the context of this survey, we argue that these vulnerabilities can only be avoided through proper understanding of the underlying Ethereum system on part of the programmer and following the best programming practices and security recommendations. For this group of vulnerabilities, Vyper or any other high level language is not a candidate to address them. However, these vulnerabilities may be addressed by design and verification tools for smart contracts~\cite{luu2016making,tsankov2018securify,brent2018vandal,nikolic2018finding,mavridou2018designing,mavridou2019verisolid,nelaturu2020verified}. For detailed discussion of such tools, we refer the reader to relevant surveys~\cite{di2019survey,kirillov2019evaluation,chen2019survey,liu2019survey,harz2018towards,parizi2018empirical}. We now proceed to describe each of these 19 vulnerabilities in Vyper's context and provide reasoning for the classification of each of these into their respective group.

\subsection{Vulnerabilities addressed by Vyper}

\subsubsection{Integer overflow and underflow}

This vulnerability occurs due to the fact that both Solidity and the EVM do not enforce integer overflow / underflow detection. This can lead to attacks which make unauthorized or unintended manipulation to a contract's state variables if proper measures were not taken during development. Libraries such as SafeMath~\cite{openzeppelin2020safemath} do provide mechanisms for protecting against over/underflows in Solidity but Vyper has this feature built-in. In Vyper, the contract execution will revert if an over/underflow is detected~\cite{feist2019watch}.

\subsubsection{DoS with unbounded operations}

This vulnerability occurs when the operations required in the execution of a function exceed the block gas limit due to unbounded operations either in the contract itself or in one of the called contracts. Vyper solves this problem by having a precise upper bound for the gas consumption of any function call. This is possible because infinite length loops and recursive function calling are not allowed in Vyper~\cite{vyper}.

\subsubsection{Unchecked call return value}

This vulnerability exists due to the discrepancy in Solidity's handling of exceptions occurring in callee contracts. Solidity handles exceptions when calling another contract in two ways: (1) when directly referencing the callee's contract instance or using the \snippet{transfer()} function; (2) when using one of the four low level methods (\snippet{call}, \snippet{staticcall}, \snippet{delegatecall}, and \snippet{send}). In the first instance, the exception is ``bubbled up'' and the entire transaction is reverted whereas in the second case only a \snippet{false} is returned to the calling contract. The uninitiated developer can be misled to think that any call(s) to other contracts were successful because no exception was thrown in the latter case. Solidity does not enforce any checks on the return values. In comparison, Vyper only provides two ways to call another contract in addition to the direct reference, i.e., the functions \snippet{send()} and \snippet{raw\_call()}~\cite{vyper_low}. The current Vyper compiler has built-in asserts for both of these functions~\cite{vyper_github}, so that in case of a failure the entire transaction will be reverted.

\subsection{Vulnerabilities partially addressed by Vyper}

\subsubsection{Reentrancy}

The reentrancy vulnerability occurs when a contract calls an external contract, handing it over the execution control, which allows the callee to call back to the calling contract and then be able to perform some malicious steps. A contract is particularly vulnerable to reentrancy attacks if it does not make the necessary state changes before calling the external contract or if the code does not protect against multi-contract access situations. Vyper provides the functionality to the programmer to protect a contract against multi-contract access situations by providing a \snippet{nonreentrant} decorator which places a lock on the current function and all functions with the same key value~\cite{vyper_non}. \cref{fig:2.1} provides an example of how to use this feature (for function \snippet{sendFunc}). If any external callee tries to callback into such functions, it will result in a revert call. Solidity did not provide such functionality and the developer had to implement locks or mutexes themselves or through some third party libraries~\cite{openzeppelin2020reentrancy}. However, even in Vyper, the developer still has to identify the functions or blocks of code that might be susceptible to such a vulnerability and also has to ensure that all necessary state changes are made before making an external interaction. The current Vyper compiler does not warn the developer for such~cases.

\begin{figure}[!htbp]
\begin{lstlisting}
sent: public(bool)

@public
def __init__():
    self.sent = False

@public
@nonreentrant("exampleKey")
def sendFunc(to: address) -> uint256:
    if self.sent == False:
        send(to, 1)
        self.sent = True
    return 1
\end{lstlisting}
\caption{Nonreentrant decorator feature.}
\label{fig:2.1}
\end{figure}

\subsection{Vulnerabilities not addressed by Vyper}

\subsubsection{Delegatecall Injection}

EVM provides the option of calling an external contract with the context of the caller contract using the \snippet{DELEGATECALL} opcode. 
This is achieved by using the \snippet{delegatecall} function in Solidity and using the \snippet{raw\_call} function with the \snippet{delegate\_call} keyword argument set to \snippet{True} in Vyper, e.g.,
\snippet{raw\_call(argAddress, example\_bytes, outsize=0, gas=10000, value=1, delegate\_call=True)}. 
However, if the contract being called is malicious, it can manipulate the state variables of caller contract. This vulnerability can be mitigated in Solidity by only using the \snippet{DELEGATECALL} with contracts that have been declared as libraries. In Vyper this vulnerability can similarly be avoided by only using \snippet{DELEGATECALL} with functions that are declared with the \snippet{@constant} decorator, which ensures that the functions will not mutate the state. However, like Solidity, this is not enforced in Vyper because there are legitimate cases, using the \snippet{DELEGATECALL} opcode, where the caller wants the callee to modify its state. Perhaps the best way to completely avoid this vulnerability is for the EVM to provide another opcode (just like \snippet{DELEGATECALL}) which retains the context of the caller contract but causes a revert if the callee tries to make any changes to the caller's state (just like \snippet{STATICCALL}). Another possible solution at the language level could be to have the compiler place appropriate checks on state variables before and after the \snippet{DELEGATECALL} opcode is used and to give the user an option to enable to disable these~checks.

\subsubsection{Forced Ether to contract}

This vulnerability occurs when the developer of the smart contract incorrectly assumes that the contracts fallback or payable function will be executed each time Ether is transferred to the contract. There are two situations in which Ether can be sent to a contract without invoking its fallback or payable functions. Firstly, when a contract that is self-destructing sends its remaining Ether to the contract or secondly, if Ether is transferred to an address even before the contract is loaded to that address. The second situation is possible because the contract addresses are deterministic and can be calculated before deploying them~\cite{solidity2018solidity}. This vulnerability is due to the design of the underlying Ethereum protocol and the developer has to be aware of Ethereum's design and functionality when writing smart contracts. This vulnerability can be avoided if the contract does not place checks on the exact values of the contract's balance (\snippet{self.balance}). Currently, the Vyper compiler does not warn the developer if checks are placed on the \snippet{self.balance} variable in the contract code.	

Another possible workaround could be to have a built-in mechanism in contracts to run the payable or fallback functions if a contract is invoked and its balance is different from its last invocation (meaning Ether was forced to the contract in between the invocations).

\subsubsection{DoS with unexpected revert}

This vulnerability occurs when an external contract causes a revert resulting in disruption of execution of the caller contract before it has completed its function~\cite{consensys_known}. The most common scenario is when the developer fails to account for the case when a payment is made to an external contract whose fallback or payable function execution results in a revert. This vulnerability is addressed in Solidity and Vyper by making use of a pull rather that push based mechanism when making external payments~\cite{ethereum_safety}. Alternatively, contracts in Solidity can take measures to handle the cases in code where an external call might throw an exception. Vyper currently, does not allow the handling of~exceptions.

\subsection{Vulnerabilities addressed in Solidity or Ethereum}

\subsubsection{Erroneous Visibility}

This vulnerability occurs when a contract's visibility is incorrectly specified and thus permits unauthorized access. Solidity used to make functions public by default if the visibility was not specified. However, this was addressed with version 0.5.0 by making it compulsory to specify visibility when defining functions~\cite{solidity_breaking}. Vyper allows functions without visibility (v0.1.0beta15) but defaults them to being of private visibility instead of public.

\subsubsection{Uninitialized storage pointer}

This vulnerability occurs due to the fact that prior to Solidity 0.5.0, if a complicated local variable (e.g., struct, array or mapping) was not explicitly initialized at the time of declaration, then the local variable's reference points to slot 0 in storage by default, possibly overwriting a state variable~\cite{solidity2018storage}. Since Solidity v0.5.0, the Solidity compiler reports an error to contracts that contain uninitialized storage pointers. Also, explicit data location (i.e., storage, memory, or calldata) for all variables of struct, array or mapping type is now mandatory in Solidity~\cite{solidity_breaking}. Vyper also mandates the initialization of local variables at the time of declaration and failure to do so results in a compile time~error.

\subsubsection{Erroneous constructor name}

This vulnerability occurred due to the fact that prior to Solidity version 0.4.22, a function declared with the same name as the contract was considered to be the contract's constructor function. A constructor function is called only once at the time of contract creation to perform initialization. If the programmer accidentally misspelled this function name then it became a public function which allowed anyone to call it and possibly compromise the contract~\cite{smart_incorrect}. This vulnerability was mitigated in Solidity version 0.4.22 by introducing the usage of mandatory keyword \snippet{constructor} when defining the constructor function~\cite{solidity_release}. Similarly, Vyper uses the keyword \snippet{\_\_init\_\_} for the constructor.

\subsection{Vulnerabilities not addressable by language or toolchain}

\subsubsection{Upgradeable Contract}

This vulnerability occurs when a contract relies on external contracts for critical functions and the external contracts can be dynamically updated~\cite{chen2019survey}. This vulnerability cannot be avoided in Vyper either. The developers have to ensure that they do not outsource critical functions to untrusted external contracts which are built such that their functionality can be dynamically updated. In the future, tools and utilities may be developed that traverse the entire call hierarchy of the contract and its callees to identify functionality which is susceptible to being updated but we are not aware of any such software available at the time of~writing.

\subsubsection{Type casts}

This vulnerability occurs due to the Solidity compiler flagging some type errors (e.g., assigning an integer value to a string type) but not all~\cite{chen2019survey}. Types are also used in direct calls, where the caller must declare the callee's interface and cast to it the callee's address when performing the call. Having some type checks may mislead the programmer to believe that all type checks are made. If the function being called doesn't exist in the callee contract then the callee contract's fallback is executed without any exception being thrown to alert the programmer. The functionality is the same in Vyper. \cref{fig:5.2} shows an example of a smart contract which defines two contract interfaces (\snippet{LibA} and \snippet{LibB}) but will have no way of knowing if the argument address passed to function \snippet{workerFunction} is of type \snippet{LibA} and not \snippet{LibB}. Smart contract development tools such as VeriSolid~\cite{nelaturu2020verified}  can be used to build and deploy smart contracts that are free from this vulnerability in Solidity.

\begin{figure}[!htbp]
\begin{lstlisting}
contract LibA:
    def featureFunc(arg: uint256): constant
    
contract LibB:
    def featureFunc(arg: uint256): constant

@public
def workerFunction(inputLib: address):
    LibA(inputLib).featureFunc(1)
\end{lstlisting}
\caption{Type-cast vulnerability in Vyper.}
\label{fig:5.2}
\end{figure}

\subsubsection{Insufficient signature information}

This vulnerability occurs in a contract that uses a signed message to authorize payments to participants (e.g., a micropayment channel contract~\cite{solidity_micro}) and the signed message can be used by the participants to claim authorization for a second action (replay attack). This vulnerability can result in replay attacks in the same contract, across multiple contracts, or even across multiple blockchains. This vulnerability was exploited for cross-blockchain replay attacks after the Ethereum classic hard fork \cite{replay} and was addressed by the Ethereum Improvement Proposal (EIP) 155~\cite{buterin2016simple}. To avoid this vulnerability within the same contract or across multiple contracts, the developer has to ensure that the contract's signed message generation and authentication mechanism is properly implemented. This can be achieved by including the requisite information (e.g., nonce and contract address) in the message~\cite{marx2018signing}. The developer can also rely on trusted standards like the ERC-721~\cite{entriken2018non} when implanting token functionality to avoid these problems.

\subsubsection{Frozen Ether}

This vulnerability is observed when Ether is stuck in a contract with no way to send it to other contracts or external accounts. This can happen due to the contract having a faulty or nonexistent function for sending Ether. It can also occur due to the contract relying on another contract for its money-spending functions and the callee contract having been deleted or not being usable anymore. Since this can occur due to a wide range of reasons we believe it is best addressed by the required due diligence on the developer's part and by not outsourcing critical spending functions to untrusted contracts. Third party development and verification tools such as VeriSolid~\cite{nelaturu2020verified} can be used to ensure that the appropriate withdrawal functions always remain reachable in the developed smart contracts.

\subsubsection{Authentication through \snippet{tx.origin}}

The \snippet{tx.origin} variable is used in Solidity as well as Vyper to refer to the original external account that initiated the transaction in question, whereas the \snippet{msg.sender} variable is used in both to refer to the sender of the message for the current call. This vulnerability occurs when an inexperienced developer mistakenly checks \snippet{tx.origin} for authentication purposes rather than \snippet{msg.sender}~\cite{vessenes2016tx}. \cref{fig:5.5} provides an example of  this vulnerability (Line 9) in function \snippet{withdrawAll}, which uses \snippet{tx.origin} to confirm the owner of the contract that is calling the function. This is an error on the developer's part due to their inadequate understanding of the Ethereum system and the Solidity / Vyper language.

\begin{figure}[!htbp]
\begin{lstlisting}
owner: address

@public
def __init__():
    self.owner = msg.sender

@public
def withdrawAll(to: address):
    assert tx.origin == self.owner
    send(to, self.balance)
\end{lstlisting}
\begin{lstlisting}
contract UserWallet:
    def withdrawAll(recipient: address): modifying

attacker: address

@public
def payFunc():
    UserWallet(msg.sender). withdrawAll(self.attacker)
\end{lstlisting}
\caption{\snippet{tx.origin} misuse in Vyper.}
\label{fig:5.5}
\end{figure}

\subsubsection{Unprotected suicide}

This vulnerability occurs due to the fact that contract bytecode and storage can be deleted from the Ethereum network by using the \snippet{SELFDESTRUCT} opcode. Both Solidity and Vyper provide functions to use this bytecode. Many contracts implement a self-destruct/suicide function. Developers have to ensure that the authentication mechanism is correctly implemented in such contracts so that only the owner and trusted third parties are able to self-destruct the contract. Developers must also ensure that their contracts do not depend on third party contracts that might be deleted in the future rendering their own contracts unusable. Development and verification tools such as VeriSolid~\cite{nelaturu2020verified} ensure that the suicide statement in smart contracts cannot be reached using an unintended execution trace.

\subsubsection{Leaking Ether to arbitrary address}

This vulnerability exists when a contract is able to send funds to a caller who is not an owner or investor or a legitimate payee of the contract. It occurs due to the contract not enforcing adequate authorization mechanisms before transferring funds or can occur as a result of the many other vulnerabilities mentioned in this survey. This vulnerability can be mitigated by the developer adapting proper authorization logic in code to ensure that only the intended recipients are able to withdraw Ether because the language is blind to the intentions of the developer.

\subsubsection{Secrecy failure}

This vulnerability occurs when developers incorrectly assume that restricting a variable / function's visibility would make its value/functionality hidden from the participants of the Ethereum network~\cite{atzei2017survey}. This is not the case due to the public nature of the blockchain. If a state variable is declared private, other contracts are not allowed to access it but participants can still see its value from transaction data. Similarly, the inner-workings of a private function are also visible to all. Hence, the vulnerability is only mitigated if the developer has an understanding of the underlying Ethereum system and cannot be addressed by the language or toolchain.

\subsubsection{Outdated compiler version}

This vulnerability occurs when a contract is compiled using an outdated compiler version which might contain unresolved bugs and vulnerabilities. This vulnerability is addressed by using the latest compiler version when compiling contracts in either Solidity or Vyper.

\section{Conclusion}
\label{sec:concl}

We presented a detailed comparison of how the known vulnerabilities that exist in the Solidity smart contract development environment translate to the Vyper development environment. We believe that most of the vulnerabilities listed in Chen et al.~\cite{chen2019survey} are either not addressable at the language / toolchain level or have already been mitigated in Solidity through its subsequent releases and do not surface in Vyper's environment. Vyper may introduce additional vulnerabilities of its own but those will only become evident once a stable version of Vyper is released and adapted by a larger development community. Based on this survey it now appears that most of the vulnerabilities that can be attributed to the language / toolchain have been addressed in Vyper, albeit at the cost of complex functionality. Most remaining vulnerabilities are either due to developers not following the recommended development practices and safety precautions or due to them having insufficient knowledge of the underlying Ethereum~system.

\paragraph*{Acknowledgements} 
The authors warmly thank Bryant Eisenbach of project Vyper for his valuable insights regarding~Vyper.
This work was supported in part by
the National Science Foundation under Grant CNS-1850510 and by NTT Research Inc.

\bibliographystyle{IEEEtran}
\bibliography{main}

\end{document}